\documentstyle{lamuphys}
\include{epsf}
\makeatletter
\let\chapter\hid@chapter
\makeatother
\begin{document}
\newcommand{\beq}{\begin{equation}}
\newcommand{\eeq}{\end{equation}}
\newcommand{\beqa}{\begin{eqnarray}}
\newcommand{\eeqa}{\end{eqnarray}}
\newcommand{\krig}[1]{\stackrel{\circ}{#1}}
\pagenumbering{arabic}
\title{Status of Three Flavor Baryon Chiral Perturbation Theory}

\author{Ulf-G.\,Mei\ss ner}

\institute{FZ J\"ulich, Institut f\"ur Kernphysik (Theorie),
D-52425 J\"ulich, Germany}

\maketitle

\vspace{-5.5cm}

\hfill KFA-IKP(TH)-1997-18

\vspace{5cm}

\begin{abstract}
I review the present status of three flavor baryon chiral perturbation
theory in the heavy fermion formalism. It is argued that precise
calculations have to include all terms quadratic in the quark masses.
As examples, I consider the chiral expansion of the octet baryon
masses, the baryon magnetic moments and kaon photoproduction off
nucleons.
\end{abstract}
\section{Introduction}

Chiral perturbation theory (CHPT) with nucleons is by now in a 
fairly mature status, see for example Bernard's talk at this
workshop. The extension to the three flavor case is important
for a variety of reasons. First, there exists a large body of interesting
phenomenology to study, like e.g. hyperon radiative and nonleptonic
weak decays, the baryon octet magnetic moments or kaon photoproduction
off nucleons. Second, and more important, are three questions which can 
not be addressed in the two flavor case:
\begin{itemize}
\item The splittings in the baryon octet and the deviations
from the octet Goldberger--Treiman relations allow to extract
information about the quark mass ratios $(m_s - \hat{m})/ (m_d - m_u)$
(with $\hat m$ the average light quark mass) and $m_s/\hat m$.
\item Precise calculations can shed light on the question
why flavor SU(3) works so well in some cases and less well in others,
which ultimately helps to understand the quark model.
\item We can study in great detail the long distance
  contributions to the nonleptonic weak interactions of the Standard
Model.
\end{itemize}
As will become clear in what follows, we are still far away from precisely
answering these questions. I will outline some recent developments based
on complete calculations to a given order in the low energy expansion
(the small momenta and meson masses underlying this expansion will be
denoted collectively by $q$ in what follows). After discussing some
technical aspects, I will focus on three different observables. This
is certainly subjective and should not be considered exhaustive. I
will also stick to the conventional scheme having only the Goldstone
boson  and the baryon octet as active degrees of freedom. For
an early review on three flavor baryon CHPT, including also
the spin--3/2 decuplet, see \cite{jmr}.

\section{Generating functional and effective Lagrangian}

The interactions of the Goldstone bosons with the ground state baryon
octet states are severely constrained by chiral symmetry. The
generating functional for Green functions of quark currents 
between single baryon states, $Z[j,\eta, \bar \eta$], is defined via
\beq
\exp  \bigl\{ i  Z[j,\eta, \bar \eta] \bigr\} 
= {\cal N} \int [dU] [dB] [d{\bar B}]  \exp 
 i \biggl[ S_M + S_{MB} + \int d^4x \langle\bar \eta B\rangle 
+\langle\bar B \eta\rangle
\biggr] \,  ,
\label{defgenfun}
\eeq
with $S_M$ and $S_{MB}$ denoting the mesonic and the meson--baryon
effective action, respectively, to be discussed below. $\eta$ and
$\bar \eta$ are fermionic sources coupled to the baryons and $j$
collectively denotes the external fields of vector ($v_\mu$), 
axial--vector ($a_\mu$), scalar ($s$) and pseudoscalar ($p$) type. 
These are coupled in the standard chiral
invariant manner. In particular, the scalar source contains the quark
mass matrix ${\cal M}$, $s(x) = {\cal M} + \ldots$. Traces in flavor
space are denoted by $\langle...\rangle$. The underlying effective Lagrangian
can be decomposed into a purely mesonic ($M$) and a meson--baryon ($MB$)
part as follows (I only consider processes with exactly one baryon in
the initial and one in the final state)
\beq
{\cal L}_{\rm eff} = {\cal L}_M + {\cal L}_{MB}
\eeq
subject to the following low--energy expansions
\beq
{\cal L}_M =  {\cal L}_M^{(2)} + {\cal L}_M^{(4)} + \ldots  \, \, ,
\quad {\cal L}_{MB} =  {\cal L}_{MB}^{(1)} + {\cal L}_{MB}^{(2)}
+ {\cal L}_{MB}^{(3)} + \ldots \,
\eeq
where the superscript denotes the chiral dimension. 
The pseudoscalar Goldstone fields ($\phi = \pi, K, \eta$) are collected in
the  $3 \times 3$ unimodular, unitary matrix $U(x)$, 
\begin{equation}
 U(\phi) = u^2 (\phi) = \exp \lbrace i \phi / F_0 \rbrace
\label{U}
\end{equation}
with $F_0$ the pseudoscalar decay constant (in the chiral limit).
Under SU(3)$_L$ $\times$SU(3)$_R$, $U(x)$ transforms as $U \to U' =
LUR^\dagger$, with $L,R \in$ SU(3)$_{L,R}$. The meson
Lagrangian with the external fields coupled in a chiral invariant
manner is standard and will not be discussed further, see \cite{gl85}.
The effective meson--baryon Lagrangian starts with terms of dimension one,
\beq
{\cal L}_{MB}^{(1)}  =  \langle\bar B \, [ \, i \nabla\!\!\!\!/ \, ,
\, B] \,\rangle \,
- \, m \langle \bar B  \, B \rangle 
+ {D \over 2} \, \langle\bar B \, \{u\!\!\!/ \gamma_5 , B \}\, \rangle
+ {F \over 2} \, \langle \bar B \, [ u\!\!\!/ \gamma_5 , B ] \,
\rangle  \,\,\, ,
\label{LMB1}
\eeq
with $m$ the average octet mass in the chiral limit.
The $3\times 3$ matrix $B$ collects the baryon octet, 
\begin{eqnarray}
B  =  \left(
\matrix  { {1\over \sqrt 2} \Sigma^0 + {1 \over \sqrt 6} \Lambda
&\Sigma^+ &  p \nonumber \\
\Sigma^-
    & -{1\over \sqrt 2} \Sigma^0 + {1 \over \sqrt 6} \Lambda & n
    \nonumber \\
\Xi^-
        &       \Xi^0 &- {2 \over \sqrt 6} \Lambda \nonumber \\} 
\!\!\!\!\!\!\!\!\!\!\!\!\!\!\!\!\! \right)  \, \, \, .
\end{eqnarray}
Under $SU(3)_L \times SU(3)_R$, $B$  transforms as any matter field,
\begin{equation} 
B \to B' = K \, B \,  K^\dagger
 \, \, \, ,
\end{equation}
with $K(U,L,R)$ the compensator field representing an element of the
conserved subgroup SU(3)$_V$. $\nabla_\mu$ denotes the covariant
derivative,
\beq
[\nabla_\mu , B] = \partial_\mu \, B + [ \, \Gamma_\mu , B \,]
\label{nablaB}
\eeq
and $\Gamma_\mu$ is the chiral connection,
\beq
\Gamma_\mu  =  \frac{1}{2}\, [ u^\dagger (\partial_\mu-ir_\mu) u +
u (\partial_\mu-il_\mu) u^\dagger ]  \,\,\, .
\eeq
Note that the first term in
Eq.(\ref{LMB1}) is of dimension one since $[ i \nabla \!\!\!\!/ \, , \,
B ] - m \, 
B = {\cal O}(q)$, \cite{gss}. The lowest order
meson--baryon Lagrangian contains two axial--vector coupling
constants, denoted by $D$ and $F$. It is important to note that to
leading order, no symmetry--breaking terms appear.
The dimension two and three terms
have been enumerated by  \cite{krause}. Treating the baryons as
relativistic spin--1/2 fields, the chiral power counting is no more
systematic due to the large mass scale $m$, $\partial_0 \, B 
\sim m \, B \sim \Lambda_\chi \, B$. This problem can be overcome in
the heavy mass formalism proposed in \cite{jm}. 
I follow here the path integral approach  developed in \cite{bkkm}.
Defining velocity--dependent spin--1/2 fields by a particular choice 
of Lorentz frame and decomposing the fields into their velocity
eigenstates (sometimes called 'light' and 'heavy' components),
\beqa
H_v (x) &=& \exp \{ i m v \cdot x \} \, P_v^+ \, B(x) \,\, , \quad
h_v (x)  =  \exp \{ i m v \cdot x \} \, P_v^- \, B(x) \,\, , \nonumber \\
P_v^\pm &=& \frac{1}{2} (1 \pm v \!\!\!/) \,\, , \,\,\, v^2 = 1 \,\,\, ,
\label{Bheavy}
\eeqa
the mass dependence is shuffled from the fermion propagator into a
string of $1/m$ suppressed interaction vertices. In this basis, the 
three flavor meson--baryon action takes the form 
\beq
S_{MB} = \int d^4x \, \biggl\{ \bar{H}_v^a \, A^{ab} H^b_v 
- \bar{h}_v^a \, C^{ab} h^b_v + \bar{h}_v^a \, B^{ab} H^b_v
+ \bar{H}_v^a \, \gamma_0 \,  {B^{ab}}^\dagger \, \gamma_0 \, h^b_v 
\biggr\}\,\,\, ,
\eeq
with $a,b= 1, \ldots ,8$ flavor indices. The 8$\times$8 matrices $A$,
$B$ and $C$ admit low energy expansions,  see \cite{mue:mei}. 
Similarly, one splits the
baryon source fields $\eta (x)$ into velocity eigenstates,
\beq
R_v (x) = \exp \{ i m v \cdot x \} \, P_v^+ \, \eta(x) \,\, , \quad
\rho_v (x) = \exp \{ i m v \cdot x \} \, P_v^- \, \eta(x) \,\, , 
\label{sourceheavy}
\eeq
and shift variables
\beq
h_v^{a'} = h_v^a - (C^{ac})^{-1} \, ( B^{cd} \, H_v^d + \rho_v^c \, )
\,\, \, ,
\eeq
so that the generating functional takes the form
\beq
\exp[iZ] = {\cal N} \, \Delta_h \, \int [dU][dH_v][d\bar{H}_v] \, \exp \bigl\{
iS_M + i S_{MB}' \, \bigr\}
\label{Zinter}
\eeq
in terms of the meson--baryon action $S_{MB}'$,
\beqa
S_{MB}' &=& \int d^4x \, \bar{H}_v^a \bigl( A^{ab} + \gamma_0
[B^{ac}]^\dagger \gamma_0 [C^{cd}]^{-1} B^{db} \, \bigr) H_v^b
\nonumber \\
&+& \bar{H}_v^a \bigl( R_v^a + \gamma_0 [B^{ac}]^\dagger \gamma_0 
[C^{cd}]^{-1} \rho_v^d \bigr) + \bigr( \bar{R}_v^a + \bar{\rho}_v^c
[C^{cb}]^{-1} B^{ba} \bigr) H_v^a \, \, .
\eeqa        
The determinant $\Delta_h$ related to the 'heavy' components is
identical to one. The generating
functional is thus entirely expressed in terms of the Goldstone bosons
and the 'light' components of the spin--1/2 fields. The action is,
however, highly non--local due to the appearance of the inverse of 
the matrix $C$. To render it local, one now expands $C^{-1}$ in powers
of $1/m$, i.e. in terms of increasing chiral dimension,
\beqa
[C^{ab}]^{-1} &=& 
\frac{\delta^{ab}}{2m} - \frac{1}{(2m)^2}\biggl\{
\langle {\lambda^a}^\dagger [i v \cdot \nabla  , \lambda^b ]\rangle
+ D \langle  {\lambda^a}^\dagger  \{S \cdot u, \lambda^b\} \rangle
\nonumber \\
&& \qquad\qquad
 + F \langle  {\lambda^a}^\dagger  [ S \cdot u, \lambda^b ] \rangle
\biggr\} + {\cal O}(q^2) \,\,\, ,
\label{Cinvexp}
\eeqa
with $S_\mu$ the Pauli--Lubanski spin vector and $u_\mu \sim i
\partial_\mu \phi/F_0 + \ldots \,$.
To any finite power in $1/m$, one can now perform the integration of
the 'light' baryon field components $N_v$ by again completing the
square,
\beqa
H_v^{a'} &=& [T^{ac}]^{-1} \, \bigl( R_v^c + \gamma_0 \, [B^{cd}]^\dagger
\, \gamma_0 \, [C^{db}]^{-1} \, \rho_v^b \, \bigr) \nonumber \\
T^{ab} &=& A^{ab} + \gamma_0 \, [B^{ac}]^\dagger \, \gamma_0 \,
[C^{cd}]^{-1} \, B^{db} \,\,\, .
\eeqa
Notice that the second term in the expression for $T^{ab}$ only starts
to contribute at chiral dimension two (and higher). In this manner,
one can construct the effective meson--baryon Lagrangian with the
added virtue that the $1/m$ corrections related to the Lorentz
invariance of the underlying relativistic theory are correctly given.
To end this section, I give the chiral dimension $D$ for processes
with exactly one baryon line running through the pertinent Feynman
diagrams,
\beq
D = 2L +1 + \sum_{d=2,4,6,\ldots} (d-2) \, N_d^M + \sum_{d=1,2,3,\ldots} 
(d-1) \, N_d^{MB} \ge 2L+1
\label{chdim}
\eeq
with $L$ denoting the number of (meson) loops, and $N_d^M$ ($N_d^{MB}$) 
counts the number of mesonic (meson--baryon) vertices of dimension $d$
(either a small momentum or meson mass).
This means that tree graphs start to contribute at order $q$
and $L$--loop graphs at order $q^{(2L+1)}$. Consequently, the low
energy constants (LECs) appearing in ${\cal L}_{MB}^{(2)}$ are all finite.

\subsection{Renormalization to third order}

Before discussing some specific examples, let
me turn to some more theoretical aspects, i.e. the problem that to one
loop in the chiral expansion divergences appear. 
The divergence structure of the one--loop generating functional to order
$q^3$ has been worked out, see \cite{mue:mei}.
It extends previous work by \cite{ecker}
for the pion--nucleon Lagrangian to the SU(3) 
case. While Ecker's method can also be used in SU(3), the fact 
that the baryons are in the adjoint representation of SU(3) whereas the
nucleons are in the fundamental representation of SU(2), complicates the
calculations considerably. In fig.~1 the various contributions to the 
one--loop generating functional together with the tree level generating 
functional at order $\hbar$ are shown. The solid (dashed) double lines 
represent the baryon (meson) propagator in the presence of external fields.
\begin{figure}[b]
\hskip 1.5in
\epsfysize=2.2cm
\epsffile{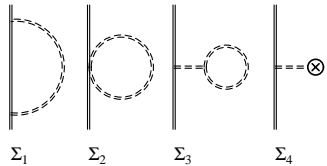}
\caption{Contributions to the one--loop generating functional 
at order $\hbar$.}
\end{figure}
Only if one ensures that the field definitions underlying the mesonic
and the baryon-meson Lagrangian match, the divergences are entirely given
by the irreducible self--energy ($\Sigma_1$) and the tadpole ($\Sigma_2$)
graphs. The explicit calculations to extract the divergences 
from $\Sigma_{1,2}$ are given in \cite{mue:mei}. 
The generating functional can be 
renormalized by introducing the following counterterm Lagrangian
\beq
{\cal L}_{MB}^{(3) \,{\rm ct}} = \frac{1}{(4\pi F_0)^2} \sum_{i=1}^{102}
\, d_i \, \bar{H}^{ab}_v (x) \, O_i^{bc}(x) \,{H}^{ca}_v (x) \,\,\, ,
\eeq
with $'a,b,c'$ SU(3) indices and the field
monomials $O_i^{bc}(x)$ are of order $q^3$. The dimensionless LECs
$d_i$ are decomposed as
\beq
d_i = d_i^r (\mu) + (4\pi)^2 \, \beta_i \, L(\mu) \,\, ,
\eeq
with
\beq
L(\mu) =\frac{\mu^{d-4}}{(4\pi)^2} \biggl\{ \frac{1}{d-4} - \frac{1}{2}
\bigl[ \log(4\pi) + 1 - \gamma \bigr]\biggr\} \,\, .
\eeq
Here, $\mu$ is the scale of dimensional regularization, $\gamma$ the 
Euler--Mascheroni constant and the $\beta_i$
are dimensionless functions of the axial couplings $F$ and $D$ that
cancel the divergences of the one--loop functional. They are tabulated
in  \cite{mue:mei} together with the $O_i^{bc}(x)$. 
These 102 terms constitute a 
complete set for the renormalization with off--shell baryons.  
As long as one is only interested in Greens functions
with on--shell baryons, the number of terms can be reduced considerably
making use of the baryon equations of motion. Also, many of these terms 
involve processes
with three or more mesons. So for calculations of kaon--nucleon scattering
or kaon photoproduction off nucleons, many of these terms do not contribute
(or only start to contribute at higher orders). At present, 
only a few of the finite $d_i^r (\mu)$ have been determined.
There are two main directions to extend these investigations. First, a 
systematic effort to pin down as many LECs as possible is needed 
and, second, the divergences at order $q^4$ should be extracted.

\subsection{The fourth order Lagrangian}

To fourth order in small momenta, the effective Lagrangian has not
been worked out in full detail. Here, I will collect the pieces
necessary to discuss the scalar sector and the baryon magnetic moments
to that order. Consider first the scalar sector.
There are three explicit symmetry breaking terms at dimension two,
\begin{equation}
{\cal L}_{MB}^{(2, {\rm br})} = b_D \, \langle\bar B \lbrace
\chi_+ , B \rbrace \rangle + b_F \, \langle\bar B [\chi_+,B]\rangle + 
b_0 \, \langle\bar B
B \rangle \, \langle \chi_+ \rangle \, \, ,   
\label{LMB2}
\end{equation}
i.e. it contains three low--energy constants, denoted $b_{0,D,F}$.
$\chi_+ = u^\dagger
\chi u^\dagger + u \chi^\dagger u$ is proportional to the quark mass
matrix ${\cal M} ={\rm diag}(m_u,m_d,m_s)$ since $\chi = 2 B_0 {\cal
  M}$. Here, $B_0 = - \langle 0 | \bar{q} q | 0 \rangle / F_0$ is
the order parameter of the spontaneous symmetry violation. 
I assume the standard scenario with $B_0 \gg F_0$.
Furthermore, one has seven independent terms contributing at dimension four,
\begin{eqnarray}
{\cal L}_{MB}^{(4)} & = & 
  e_1 \, \langle \bar{B} [\chi_+ , [ \chi_+ , B]] \rangle   
 + e_2 \, \langle\bar{B} [\chi_+ , \{ \chi_+ , B\} ] \rangle \nonumber \\   
& + & e_3 \,\langle \bar{B} \{ \chi_+ , \{ \chi_+ , B\} \} \rangle   
 + e_4 \, \langle\bar{B} \chi_+ \rangle \langle \chi_+ B\rangle \nonumber \\
& + & e_5 \, \langle \bar{B}  [ \chi_+ , B]\rangle \langle\chi_+\rangle   
 + e_6 \, \langle\bar{B}B\rangle \langle \chi_+\rangle
 \langle\chi_+ \rangle 
+ e_7 \, \langle\bar{B}B\rangle \langle \chi_+^2 \rangle \, \, \, ,
\label{leff4}
\end{eqnarray}
making  use of the Cayley--Hamilton identity.
The $e_i$ have dimension [mass$^{-3}$]. It is important to note that
some of the $e_i$ simply amount to quark mass renormalizations of some
of the dimension two LECs. To be specific, one can absorb
the effects of $e_5$ and $e_6$ in $b_F$ and $b_0$, respectively, as 
follows
\begin{equation}
b_F \to b_F - e_5 \, \langle \chi_+ \rangle \, \, ,\quad
b_0 \to b_0 - e_6 \, \langle \chi_+ \rangle \, \, .
\label{qmrenor}
\end{equation}
This is a very general phenomenon of CHPT calculations in higher
orders. For example, in $\pi \pi$ scattering there are six LECs at two
loop order $(q^6)$, but only two new independent terms
$\sim s^3$ and $\sim s \, M_\pi^4$. The other four LECs make the
$q^4$ counter terms $\bar{\ell}_i$ ($i=1,2,3,4)$ quark mass dependent. At
this point, one has two options. One can either treat the higher order
LECs as independent from the lower order ones or lump them together to
mimimize the number of independent terms. In the latter case, one
needs to refit the numerical values of the lower dimension LECs. If
one uses e.g. resonance saturation to estimate the LECs, one should
work with the first option and treat all the $e_i$
separately from the $b_i$. Let me now turn to the magnetic moments,
i.e. construct the terms involving the electromagnetic field strength
tensor. First, I need the pertinent terms of the lowest
order chiral meson--baryon Lagrangian of dimension two,
\begin{equation} \label{LMB2mm}
{\cal L}_{MB}^{(2)} = 
-\frac{i}{4m} \, b_6^F \, \langle \bar{B} [S^\mu,
S^\nu][ F_{\mu \nu}^+, B] \rangle   
-\frac{i}{4m} \,b_6^D \, \langle \bar{B} [S^\mu,
S^\nu]\{ F_{\mu \nu}^+, B\} \rangle   
\end{equation}
with $F_{\mu \nu}^\dagger = -e(u^\dagger Q F_{\mu\nu}u+
uQF_{\mu\nu}u^\dagger )$,
$F_{\mu \nu}$ the conventional photon field strength tensor and
$Q={\rm diag}(2,-1,-1)/3$ the quark charge matrix.
It is straightforward to construct the terms contributing to the counterterm
(tree) contributions with exactly one insertion from the dimension
four effective Lagrangian. For simplicity, I consider the ones related
to the explicit breaking of SU(3) due to the large
strange quark mass. These have the form, \cite{bos}:
\begin{eqnarray}\label{LMB4}
\nonumber
{\cal L}^{(4)}_{MB}  = 
& - & \frac{i\alpha_1}{4m} 
\langle \bar{B}[S^\mu,S^\nu]\left[[F^+_{\mu\nu},B],\chi^+\right] \rangle
-\: \frac{i\alpha_2}{4m} 
\langle \bar{B}[S^\mu,S^\nu]\left\{[F^+_{\mu\nu},B],\chi^+\right\}
\rangle \\
\nonumber
& - & \frac{i\alpha_3}{4m} 
\langle \bar{B}[S^\mu,S^\nu]\left[\{F^+_{\mu\nu},B\},\chi^+\right] \rangle
-\: \frac{i\alpha_4}{4m} 
\langle \bar{B}[S^\mu,S^\nu]\left\{\{F^+_{\mu\nu},B\},\chi^+\right\} \rangle
\\
& - &\frac{i\beta_1}{4m} 
\langle \bar{B}[S^\mu,S^\nu]B \rangle \langle \chi^+ F^+_{\mu\nu} \rangle
\:.
\end{eqnarray}
Here, $\chi^+$ is the spurion,
$\chi^+ = \,$diag(0,0,1), i.e. a factor $m_s$ has been
pulled out and absorbed in the low--energy constants $\alpha_{1,2,3,4}$ 
and $\beta_1$. The terms given in eq.(\ref{LMB4}) are of chiral dimension
four since $m_s = {\cal O}(q^2)$ and $F_{\mu \nu}  = {\cal
  O}(q^2)$.  Of course, in general these five terms
should be written in terms of the full quark matrix, but since $m_s
\gg m_d, m_u$, it is legitimate to neglect at this order the pionic 
contribution.  There are two more terms which could
contribute. These have the form 
\begin{equation} \label{LMB4mq}
{\cal L}_{MB}^{(4')} = 
-\frac{i \tilde{b}_6^F}{4m} \, \langle \chi_+ \rangle \,
\langle \bar{B} [S^\mu,
S^\nu][ F_{\mu \nu}^+, B] \rangle   
-\frac{i \tilde{b}_6^D}{4m} \, \langle \chi_+ \rangle \,
\langle \bar{B} [S^\mu,
S^\nu]\{ F_{\mu \nu}^+, B\} \rangle \,\, .  
\end{equation}
As before, these two terms  obviously
amount to quark mass renormalizations of $b_6^{D,F}$, i.e.
$b_6^{D,F} \to  b_6^{D,F} + \langle \chi_+ \rangle
 \, \tilde{b}_6^{D,F}$.
Their contribution can therefore be absorbed in the values of the 
corresponding dimension two LECs. We note that the seven terms given
in eqs.(\ref{LMB4},\ref{LMB4mq}) have already been enumerated in 
\cite{jlms} (in other linear combinations). This is all the
machninery needed for what follows.

\section{Kaon Photoproduction}

\noindent Pion photo-- and electroproduction in the threshold
region has been studied intensively  over the last few years by Bernard, 
Kaiser and myself (see Bernard's contribution to these proceedings) with
high precision data coming from MAMI, SAL and NIKHEF. In addition,
at the electron stretcher ring ELSA (Bonn) ample kaon photoproduction data
have been taken over a wide energy range. Only a small fraction of these
data is published in \cite{bock}, 
the larger fraction is still in the process of being
analyzed. It therefore seems timely to study the reactions $\gamma p \to
\Sigma^+ K^0$, $\Lambda K^0$ and $\Sigma^0 K^+$ in the framework of CHPT. 
This has been done in an exploratory study published in \cite{ste:mei}. 
It only has been done to third order in small momenta and thus it is
obvious that one can not expect a high precision. However, before
embarking on a full scale $q^4$ calculation, one first has to find
out whether the strict CHPT approach is at all applicable.
Here, I will critically summarize the status
of these calculations, some more details are given by Steininger in
these proceedings. 
 The threshold energies for these three processes are
$1046$, $1048$ and $911\,$MeV, in order. In the threshold region, for energies
less than $100\,$MeV above the respective threshold, it is advantageous to
perform a multipole decomposition. It suffices to work with S-- and P--wave 
multipoles.
The transition current for the process $\gamma^* (k) +
p(p_1) \to M(q) + B(p_2)$ ($M = K^+, K^0$, $B = \Lambda, \Sigma^0, \Sigma^+$)
calculated to ${\cal O}(q^3)$ can be decomposed into Born, one--loop 
and counterterm contributions,
\beq
T = T^{\rm Born} + T^{\rm 1-loop} + T^{\rm c.t.} \quad ,
\eeq
where the Born terms subsume the leading electric and the subleading
magnetic couplings of the photon to the nucleon/hyperon and $\gamma^*$
denotes a real ($k^2 = 0$) or virtual ($k^2 < 0$) photon. The calculation
of the Born terms is standard, for charged kaon production the SU(3) 
generalization of the Kroll--Rudermann term gives the dominant contribution
to the electric dipole amplitude. Of particular interest is the observation
that the leading P--wave multipoles for $\Sigma^0 K^+$ production
are very sensitive
to the yet unmeasured magnetic moment of the $\Sigma^0$ because it is enhanced
by the coupling constant ratio $g_{p K \Lambda} / g_{p K {\Sigma^0}} = (D+3F)/
\sqrt{3} / (F-D) \simeq -5$. The one loop graphs are also easy to calculate.
Two remarks concerning these are in order. First, the SU(3) calculation allows
one to investigate the effect of kaon loops on the SU(2) predictions.
As expected, it is found that these effects are small, e.g. for neutral
pion photoproduction off protons,
$E_{0+, {\rm thr}}^K = (e F M_\pi^3 )/( 96\pi^2 F_\pi^3 M_K) =
0.14 \cdot 10^{-3}/ M_{\pi^+}$,
which is just 1/10th of the empirical value and considerably smaller than
the pion loop contribution. This result  is in agreement with
the famous decoupling theorem. In the chiral SU(2) limit, that is for a fixed
strange quark mass, kaon loop effects must decouple, which means that they
are suppressed by inverse powers or logs of the heavy mass, here $ M_K$. 
Second, the loop graphs give rise to the imaginary part of the transistion
amplitude. Here, one encounters the standard problem of CHPT, namely that
at a given order the imaginary parts are given to much less precision than
the real ones due to the ${\cal O}(p^{2N})$ suppression for $N$--loop graphs. 
One finds that these imaginary parts come out much too big, which is caused
by the pion loops.  This can be understood by considering the rescattering 
graph $\gamma p  \to \pi^+ n \to Y K^+$. By virtue of the Fermi--Watson
theorem, one finds
${\rm Im}~E_{0+} = {\rm Re}~E_{0+}^{\pi^+ n} \cdot a_{\pi K} \cdot
{\rm PS}$,
where PS denotes the phase space allowed for the virtual pion and $a_{\pi K}$
the $\pi K$ scattering length in the respective channel. Obviously, the
initial charged pion photoproduction process is far away from its threshold,
out of the range of applicability of CHPT. In \cite{ste:mei} 
these multipoles were 
thus taken from the existing multipole analysis. Clearly, this
needs refinement. At next order in the chiral expansion, one has e.g. 
additional contributions from $\pi^0$ and $\eta$ rescattering graphs. Note also
that to this order, $q^3$, the loop graphs are not finite but need standard
renormalization. This can either be done by direct Feynman diagram calculation
or by using the general method described above (this particular 
example is worked out in detail in \cite{mue:mei}). 
Finally, there are the counter 
terms.  Altogether, there are 13 various operators with unknown
low--energy constants. One combination, $d_1+d_2$, can be fixed from
the nucleon axial radius. This also constrains the yet unmeasured
$p \to \Lambda\, K^+$ transition axial radius,
\beq
\langle r_A^2 \rangle_{p \to \Lambda \, K^+} = \frac{3
 \sqrt{2}}{D+3F} (d_1 + 3 d_2) = 0.23 \ldots 0.70 \,\, {\rm fm}^2 \,\, ,
\eeq
To fix the other LECs, $d_3 , \ldots , d_{13}$,  resonance saturation
including the baryon decuplet and the vector meson nonet was used.
I now summarize the results for the various
final states (photoproduction case).

\noindent ${\underline{K^0 \Sigma^+}:}$ All LECs are determined by
resonance exchange. The total cross section has been calculated for the
first 100 MeV above threshold. No  data point exists in this range so
far, but soon the new ELSA data should be available. 
The electric dipole amplitude is real at threshold, we
have $E_{0+}^{\rm thr} (K^0 \Sigma^+) = 1.07 \times 10^{-3}/M_{\pi}$.
We also have given a prediction for the recoil polarization at
$E_\gamma = 1.26\,$GeV (which is the central energy of the lowest bin
of the not yet published ELSA data). 

\noindent ${\underline{K^+ \Lambda}:}$The total cross section from
threshold up to 100 MeV above is shown in Fig.2a (left panel). The lowest
bin from ELSA, see \cite{bock},  $E_\gamma \in [0.96,1.01]\,$GeV,  has
$\sigma_{\rm tot}~=~(1.43 \pm 0.14)\,~\mu$b, i.e. we slightly
underestimate the total cross section. In Fig.2b (left), I show the predicted
recoil polarization $P$ at $E_\gamma = 1.21\,$ GeV (which is higher in
energy than our approach is suited for). Amazingly, the shape and
magnitude of the data is well described for forward angles,
but comes out on the small side for backward angles.  
Most isobar models,
that give a descent description of the total and differential 
cross sections also at higher energies, fail to explain this angular
dependence of the recoil polarization. 

\noindent ${\underline{K^+ \Sigma^0}:}$The total cross section is
shown in Fig.2a (right panel). It agrees  with the two data points from 
ELSA. The recoil polarization at $E_\gamma =
1.26\,$GeV is shown in Fig.2b (right). It has the right shape but comes out
too small in magnitude. Nevertheless, one observes the important sign 
difference to the $K^+ \Lambda$ case, which is commonly attributed to 
the different quark spin structure of the $\Lambda$ and the 
$\Sigma^0$. Notice that this argument is strictly correct
for massless quarks only.
Here, it stems from an intricate interference of the complex S-- and
P--wave multipoles. 
In any case, one would like to have data closer to
threshold and with finer energy binning to really test the CHPT scheme.
\begin{figure}[t]
\vspace{9.cm}
\hspace{-1.5cm}
\includegraphics{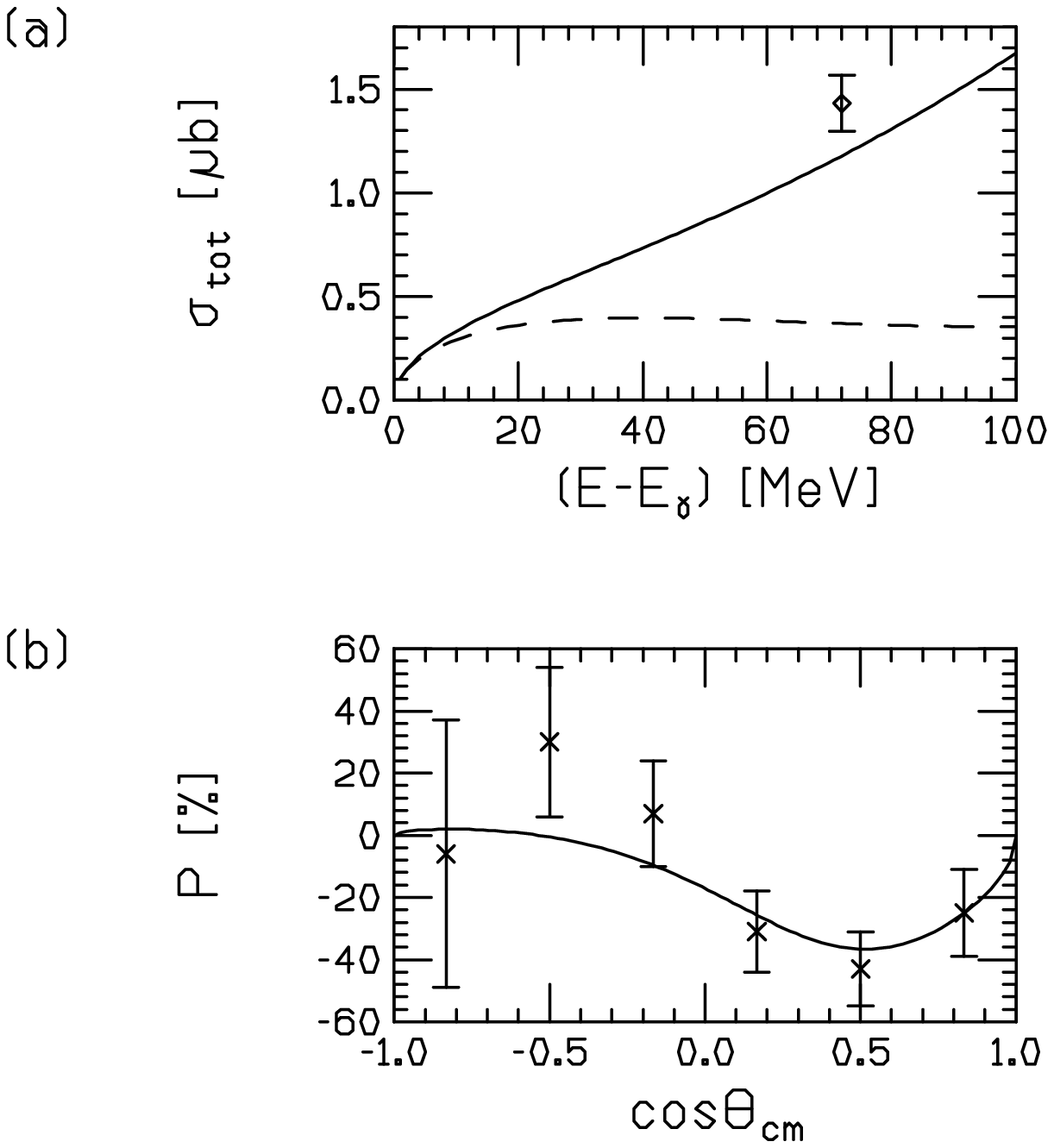}
\hspace{5.5cm}
\includegraphics{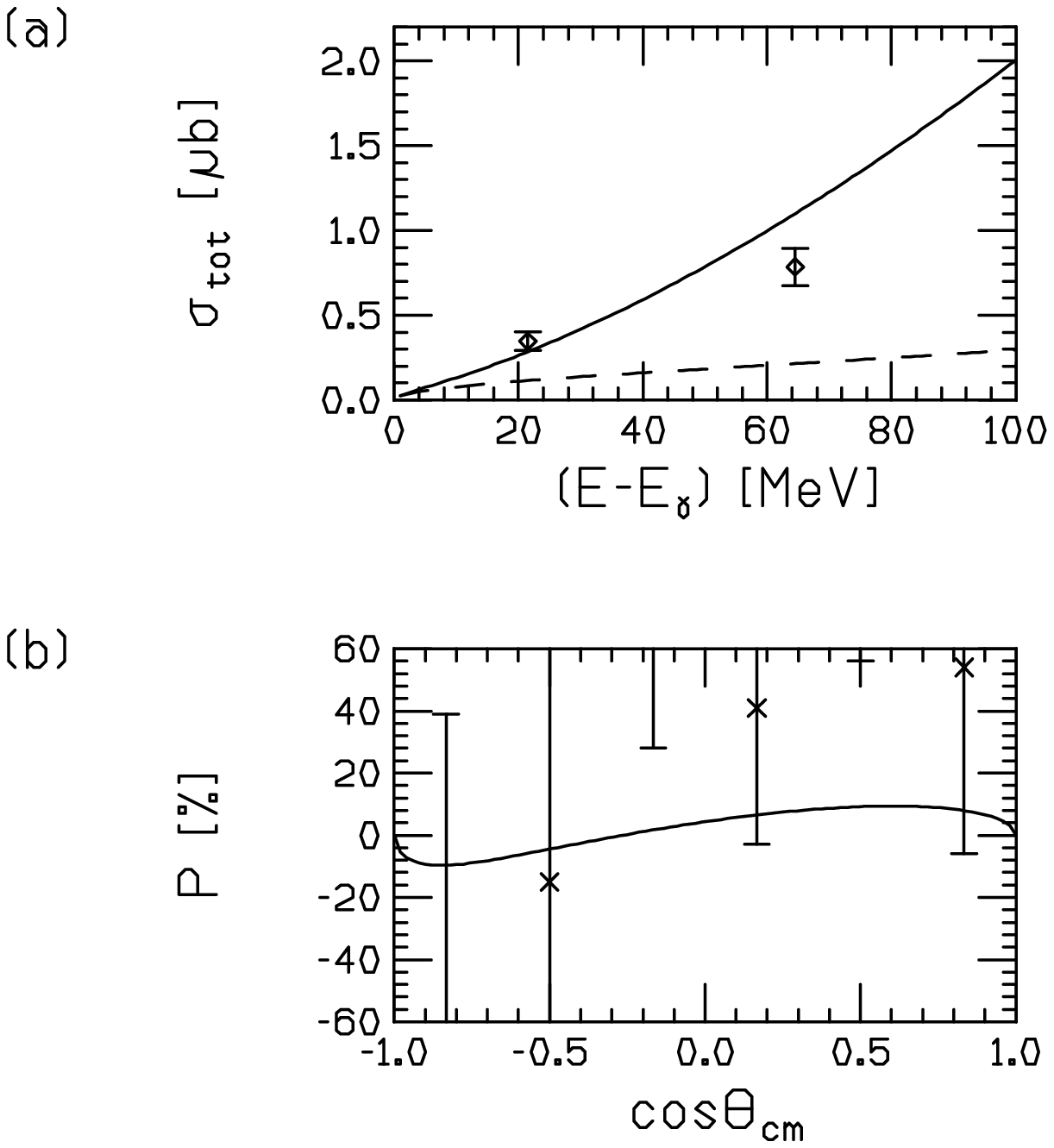}
\vspace{-1.5cm}
\caption{
Left panel: (a) Total cross section for $\gamma p \to K^+ \Lambda$
(solid line). The S--wave contribution is given by the dashed line.
(b) Recoil polarization at $E_\gamma = 1.21 \,$GeV. 
Right panel: (a) Total cross section for $\gamma p \to K^+ \Sigma^0$.
(b) Recoil polarization at $E_\gamma = 1.26 \,$GeV. 
The data are from Bockhorst et al. (1994).}
\end{figure}
Clearly, these results 
should only be considered indicative since one should include (a) higher 
order effects (for both the S-- and P--waves), (b) higher partial waves
and (c) has to  get a better handle
on the ranges of the various coupling constants. In addition, one would
also need more data closer to threshold, i.e. in a region where the
method is applicable. However, the results presented are encouraging 
enough  to pursue a more detailed study of these reactions (for real 
and virtual photons) in the framework of "strict" chiral perturbation theory.

\section{Baryon Masses}
The scalar sector of baryon CHPT is particularly interesting since
it is sensitive to scalar--isoscalar operators and thus directly to the
symmetry breaking of QCD. This is most obvious for the pion-- and 
kaon--nucleon $\sigma$--terms, which measure the strength of the scalar
quark condensates $\bar q q$ in the proton. Here, $q$ is a generic symbol for 
any one of the light quarks $u$, $d$ and $s$. Furthermore, the quark mass
expansion of the baryon masses allows to gives bounds on the ratios of the
light quark masses, see \cite{juerg}. 
The most general effective Lagrangian to fourth 
order   necessary to investigate the scalar sector 
consists of seven dimension two and seven dimension four terms
with LECs plus some additional dimension two terms with fixed 
coefficients $\sim 1/m$. The dimension two terms with LECs fall into 
two classes, one related to symmetry breaking and the other are 
double--derivative meson-baryon vertices. The LECs related to the latter 
ones can be estimated with some confidence from resonance exchange. 
A method to estimate the symmetry breakers will be discussed below. 
Although the analysis of the octet baryon masses in the framework of chiral
perturbation theory already has a long history, only
recently the results of a  calculation including  all terms of second 
order in the light quark masses, ${\cal O}(m_q^2)$, were presented in
\cite{bo:mei}. 
The calculations were performed  in the isospin limit $m_u = m_d$ and the 
electromagnetic corrections were neglected. Previous investigations considered
mostly the so--called computable corrections of order $m_q^2$ or included 
some of the finite terms at this order. This, however, contradicts the spirit 
of CHPT in that all terms at a given order have to be retained. 
The quark mass expansion of the octet baryon masses takes the form
\begin{equation}
m = \, \, \krig{m} + \sum_q \, B_q \, m_q + \sum_q \, C_q \, m_q^{3/2} + 
\sum_q \, D_q \, m_q^2  + \ldots
\label{massform}
\end{equation}
modulo logs. Here, $\krig{m}$ is the octet mass in the chiral limit of
vanishing quark masses and the coefficients $B_q, C_q, D_q$ are 
state--dependent. Furthermore, they include contributions proportional
to some LECs which appear beyond leading order in the effective Lagrangian.
In contrast to the ${\cal O}(q^3)$ calculation, which gives the leading
non-analytic terms $\sim m_q^{3/2}$, the order $q^4$ one is no longer finite
and thus needs renormalization. Intimately connected to the baryon masses
are the $\sigma$--terms (I only consider $\sigma_{\pi N}$ on what
follows),
\begin{equation}
\label{defsigma}
\sigma_{\pi N} (t)  =  \hat m \, \langle p' \, 
| \bar u u + \bar d d| \, p \rangle
\, \, \, , 
\end{equation}
with $|p\rangle$ a proton state with four--momentum $p$ and $t 
= (p'-p)^2$ the 
invariant momentum transfer squared. A relation between $\sigma_{\pi N} (0)$
and the nucleon mass is provided by  the Feynman--Hellmann theorem, 
$\hat{m}(\partial m_N / \partial \hat{m}) = \sigma_{\pi N} (0)$, with 
$\hat m$ the average light quark mass. Furthermore, the strangeness fraction 
$y$ and $\hat \sigma$ are defined via
\begin{equation}
y = \frac{ 2 \, \langle p| \bar  s s|p \rangle}
{\langle p|\bar u u + \bar d d |p \rangle} \equiv 1 -
\frac{\hat \sigma}{\sigma_{\pi N} (0)} \,\, .
\label{defy}
\end{equation}
I now discuss some of the results presented in~\cite{bo:mei}. Some
more details are given by Borasoy in these proceedings. As stated before, there
are ten LECs related to symmetry breaking. Since there do not exist enough 
data to fix all these, they were estimated  by means of resonance exchange. 
To deal with such scalar-isoscalar operators, the standard resonance 
saturation picture based on tree graphs was  extended to include loop diagrams.
In contrast to the two-flavor case, the scalar mesons in SU(3) can not explain 
the strength of the symmetry breakers because these mesons
 are not effective degrees
of freedom parametrizing strong pionic/kaonic correlations. To be precise,
the dimension two symmetry breakers can be estimated by performing a 
best fit to the baryon masses based on a ${\cal O}(q^3)$ calculation,
see \cite{bkmz}. For scalar couplings of ``natural'' size, 
these values can not  even be reproduced within one order of magnitude.
One way to solve this problem, although it has its own conceptual
difficulties, is to consider besides standard tree graphs with scalar meson 
exchange also Goldstone boson loops with intermediate baryon resonances 
(spin--3/2 decuplet and the spin--1/2 (Roper) octet)
for the scalar--isoscalar LECs. In \cite{bo:mei} a consistent scheme 
to implement 
resonance exchange under such circumstances was developed. In particular, it
avoids double--counting and abids to the strictures from analyticity. Within 
the one--loop approximation and to leading order in the resonance masses, the 
analytic pieces of the pertinent graphs are still divergent, i.e.
one is left with three a priori undetermined renormalization constants
($\beta_\Delta$, $\delta_\Delta$ and $\beta_R$). These have to be
determined together with the finite scalar meson--baryon couplings $F_S$ 
and $D_S$ and the octet mass in the chiral limit. Using the baryon masses and
the value of $\sigma_{\pi N} (0)$ as input, one can determine all LECs
in terms of one parameter, $\beta_R$. This parameter can be shown to be 
bounded and the  observables are insensitive to variations of it within 
its allowed range. Furthermore, it was also demonstrated
that the effects of two (and higher) loop diagrams can almost entirely
be absorbed in a redefinition of the one loop renormalization parameters. 
Within this scheme, one finds for the octet baryon mass in the chiral limit
$\krig{m} = 770\pm 110\, {\rm MeV}$. The quark mass expansion of the baryon 
masses,  in the notation of Eq.(\ref{massform}), reads
\begin{eqnarray}
&& m_N  = \,  \krig{m}  \, ( 1 + 0.34 - 0.35 + 0.24 \, ) \, \, ,
\nonumber \\
&& m_\Lambda  = \,  \krig{m} \, ( 1 + 0.69 - 0.77 + 0.54 \, ) \, \, ,
\nonumber \\
&& m_\Sigma  = \,  \krig{m} \, ( 1 + 0.81 - 0.70 + 0.44 \, ) \, \, ,
\nonumber \\
&& m_\Xi  = \,  \krig{m} \, ( 1 + 1.10 - 1.16 + 0.78 \, ) \, \, .   
\label{mexpand}
\end{eqnarray}
One observes that there are large cancellations between the second order
and the leading non--analytic terms of order $q^3$, a well--known effect.
The fourth order contribution to the nucleon mass is fairly small, whereas it
is sizeable for the $\Lambda$, the $\Sigma$ and the $\Xi$. This is
partly due to the small value of $\krig{m}$, e.g. for the $\Xi$ the
leading term in the quark mass expansion gives only about 55\% of the
physical mass and the second and third order terms cancel almost completely.
From the chiral expansions exhibited in Eq.(\ref{mexpand}) one can not    
yet draw a final conclusion about the rate of convergence in the
three--flavor sector of baryon CHPT. Certainly, the breakdown of CHPT
claimed by \cite{juerg} is not observed. On the other hand, the
conjecture by \cite{jm2} that only the leading non--analytic corrections (LNAC)
$\sim m_q^{3/2}$ are large and that further terms like the ones $\sim m_q^2$
are moderately small, of the order of 100 MeV, is not supported. 
The chiral expansion of the $\pi N$ $\sigma$--term shows a moderate 
convergence, i.e. the terms of increasing order become successively smaller,
\begin{equation}
\sigma_{\pi N} (0) = 58.3 \, ( 1 - 0.56  + 0.33) \, \, \, {\rm MeV} 
= 45 \, \, {\rm MeV}     \, \, .
\label{signo}
\end{equation}
Still, the $q^4$ contribution is important.  For the strangeness fraction  
$y$ and $\hat \sigma$, one finds
\begin{equation}
y = 0.21 \pm 0.20 \, \, , \, \,
\hat \sigma = 36 \pm 7 \, \, {\rm MeV} \, \, .
\label{valunc}
\end{equation}
The value for $\hat \sigma$ compares favourably
with Gasser's estimate, $\hat \sigma = 33 \pm 5\,$MeV. 
Finally, a
comment concerning the difference of the pion--nucleon $\sigma$--term 
at $t=0$ and at the Cheng--Dashen point is in order. In \cite{bkmcd}
it was shown that the remainder $\Delta_R$ not fixed by chiral symmetry, 
i.e. the
difference between the on--shell $\pi N$ scattering amplitude $\bar{D}^+
(0,2M_\pi^2)$ and the scalar form factor $\sigma_{\pi N} (2M_\pi^2)$,
contains no chiral logarithms and vanishes simply as $M_\pi^4$ in the
chiral limit. In addition, an upper limit was reported, 
$\Delta_R \le 2\,$MeV. While this is a small effect, it is an
important ingredient in the analysis of the $\sigma$--term.

\section{Magnetic Moments} 
The magnetic moments of the octet baryons  have been measured with
high precision over the last decade. On the theoretical side, SU(3)
flavor symmetry was first used by \cite{CG} to  
predict seven relations between the eight moments of the $p$, $n$,$\Lambda$,
$\Sigma^\pm$, $\Sigma^0$, $\Xi^-$, $\Xi^0$ and the $\mu_{\Lambda \Sigma^0}$ 
transition moment in terms of two parameters.  One of these relations is
in fact a consequence of isospin symmetry alone. In modern language, 
this was a tree level calculation with the lowest order effective
chiral meson--baryon Lagrangian of dimension two  given in
Eq.(\ref{LMB2mm}) and depicted in fig.3a.
Given the simplicity of this approach, these relations work remarkably well,
truely a benchmark success of SU(3).
\begin{figure}[t]
\hskip 1.4in
\epsfxsize=1.9in
\epsffile{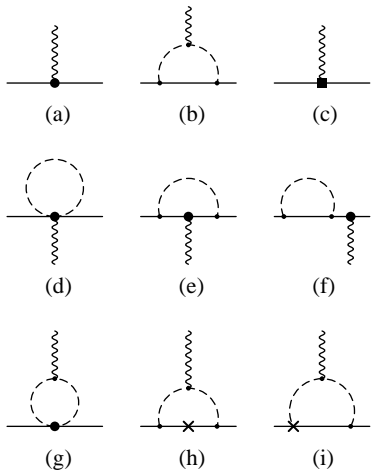}
\caption{Chiral expansion of the magnetic moments
to order $q^2$ (a), $q^3$ (b) and $q^4$ (c-i).
The cross denotes either a $1/m$ insertion with a fixed LEC or
one related to scalar symmetry breaking. The dot is
the leading dimension two insertion and the box depicts the
counterterms at fourth order.} 
\end{figure}
The first loop corrections arise at order $q^3$ in 
the chiral counting, see \cite{CP} (cf. fig.3b). They are given entirely in 
terms of the lowest order parameters from the dimension one (two)
meson--baryon (meson) Lagrangian. It was
found that these loop corrections are large for standard values of the
two axial couplings $F$ and $D$.  \cite{CP} derived
three relations independent of these coupling constants.
These are, in fact, in good agreement with the data. However, the deviations
from the Coleman--Glashow relations get considerably worse. This fact has
some times been taken as an indication for the breakdown of SU(3) CHPT.
To draw any such conclusion, a calculation of order $q^4$ is mandatory.
This was attempted by \cite{jlms}, however, not all terms were accounted for.
To be precise, in that calculation the contribution from the
graphs corresponding to fig.3c-f were worked out. As pointed out 
by \cite{mei:ste},
there are additional one--loop graphs at ${\cal O}{(q^4)}$, namely tadpole
graphs with double--derivative meson--baryon vertices (fig.~3g) and diagrams
with fixed $1/m$ and symmetry breaking insertions $\sim b_{D,F}$
from the dimension two Lagrangian, see
fig.~3h,i. Some (but not all) of these contributions were implicitely
contained in the work of \cite{jlms} as becomes obvious
when one expands the graphs with intermediate decuplet states. 
However, apart form these decuplet contributions to the LECs (in the
language used here), there are also important t--channel vector meson
exchanges which are not accounted for if one includes the spin--3/2
decuplet in the effective theory and calculates the corresponding
tree and loop graphs. In total, there
are seven LECs related to symmetry breaking and three related to scattering
processes (the once appearing in the class of graphs fig.~3g). 
These latter LECs
can be estimated with some accuracy from resonance exchange. The strategy
of \cite{mei:ste}
was to leave the others as free parameters and fit the magnetic
moments. One is thus able to investigate the chiral expansion of the
magnetic moments and to predict the ${\Lambda \Sigma^0}$ transition moment.
The chiral expansion of the various magnetic moments thus takes the
form
\beq 
\label{mubform}
\mu_B = \mu_B^{(2)} + \mu_B^{(3)} + \mu_B^{(4)}
= \mu_B^{(2)} \,( \, 1 + \varepsilon^{(3)} + \varepsilon^{(4)} \, ) \quad ,
\eeq
with the result (all numbers in nuclear magnetons)
\begin{eqnarray} \label{conv}
\begin{array}{llrllr}
\mu_p                 & = & 4.48  & (1 - 0.49 + 0.11) & = & 2.79 \, , \\
\mu_n                 & = & -2.47 & (1 - 0.34 + 0.12) & = & -1.91 \, , \\
\mu_{\Sigma^+}        & = & 4.48  & (1 - 0.62 + 0.17) & = & 2.46 \, , \\
\mu_{\Sigma^-}        & = & -2.01 & (1 - 0.31 - 0.11) & = & -1.16 \, , \\
\mu_{\Sigma^0}        & = & 1.24  & (1 - 0.87 + 0.40) & = & 0.65 \, , \\
\mu_\Lambda           & = & -1.24 & (1 - 0.87 + 0.37) & = & -0.61 \, , \\
\mu_{\Xi^0}           & = & -2.47 & (1 - 0.89 + 0.40) & = & -1.25 \, , \\
\mu_{\Xi^-}           & = & -2.01 & (1 - 0.64 - 0.03) & = & -0.65 \, , \\
\mu_{\Lambda\Sigma^0} & = & 2.14  & (1 - 0.53 + 0.19) & = & 1.40 \, , \\
\end{array} 
\end{eqnarray}
setting here the scale of dimensional regularization $\lambda =
800\,$ MeV.
In all cases the ${\cal O}(q^4)$ contribution is smaller than the one 
from order $q^3$ by at least a factor of two, in most cases even by 
a factor of three. 
Like in the case of the baryon masses, one finds sizeable
cancellations between the leading and next--to--leading order terms
making a {\it precise} calculation of the ${\cal O}(q^4)$ terms absolutely
necessary. In particular, the previously omitted terms with a
symmetry-breaking insertion from Eq.(\ref{LMB2}) (fig.3h)
turn out to be very important. One can
predict the transition moment to be $\mu_{\Lambda \Sigma^0} 
= (1.40 \pm 0.01) \mu_N$ in agreement with the lattice gauge theory 
result of \cite{lwd}, $\mu_{\Lambda \Sigma^0} = (1.54 \pm 0.09) \mu_N$. 
\cite{jlms} derived one relation amongst the magnetic moments
independent of the axial couplings. This relation does not hold any
more in the complete ${\cal O}(q^4)$ calculation. To be specific, the
graphs fig.3h with a scalar symmetry breaking insertion do not respect
this relation (for details, consult the appendix in \cite{mei:ste}).
Of course, this is not quite the end of the story. What is certainly missing
is a deeper understanding of the numerical values of the symmetry breaking 
LECs which were used as fit parameters by \cite{mei:ste}.

\section{Summary and outlook}

A final conclusion on the convergence of three flavor baryon chiral
perturbation theory can not yet be drawn. While the scalar sector 
still looks somewhat discouraging (which might be due to the model
used to estimate the LECs), the first steps including electromagnetic
probes like for the magnetic moments and kaon photoproduction off 
protons appear encouraging. From the long list of topics to be
addressed I personally view three as most relevant: a) we should try
to sharpen the calculations of the mass splittings in the baryon
octet, b) repeat the analysis of the Goldberger--Treiman discrepancies
(GTDs) and c) finally settle the old problem of the hyperon nonleptonic
decays first addressed by \cite{bsw}. Clearly, more precise data
are mandatory, for example the present day knowledge on the $NYK$
coupling constants is absolutely insufficient for extracting stringent
bounds on $m_s/\hat m$ from the octet GTDs. I am hopeful that more 
and more precise data will come from ELSA, Jefferson Lab, MAMI 
and DA$\Phi$NE helping us to resolve all these puzzles.

%

%
%


\begin{thebibliography}
%
\bibitem{}{bkkm}{Bernard et al. (1992)}
Bernard,\,V., Kaiser,\,N., Kambor,\,J., Mei\ss ner,\,Ulf-G. (1992):
Chiral structure of the nucleon.
Nucl.\,Phys. {\bf B388},  315-345
%
\bibitem{}{bkmz} {Bernard et al. (1993)}
Bernard,\,V.,  Kaiser,\,N.,  Mei\ss ner,\,Ulf-G. (1993):
Critical analysis of baryon masses and sigma--terms in heavy baryon
chiral perturbation theory. 
Z. Phys. {\bf C60}, 111-119.
%
\bibitem{}{bkmcd} {Bernard et al. (1997)}
Bernard,\,V.,  Kaiser,\,N.,  Mei\ss ner,\,Ulf-G. (1997):
On the analysis of the pion-nucleon $\sigma$-term: The size of the
remainder at the Cheng-Dashen point.
Phys. Lett. {\bf B389}, 144-148
%
\bibitem{}{bsw} {Bijnens et al. (1985)}
Bijnens,\,J., Sonoda,\,H., Wise,\,M.B. (1985):
On the validity of chiral perturbation theory for weak hyperon decays.
Nucl. Phys. {\bf B261}, 185-198
\bibitem{}{bock} {Bockhorst et al. (1994)} 
Bockhorst,\,M. et al. (1993): Measurement of $\gamma p \to K^+
\Lambda$ and $\gamma p \to K^+ \Sigma^0$ at photon energies up
to 1.47 GeV. Z. Phys. {\bf C63}, 37-47
%
\bibitem{}{bo:mei}{Borasoy and Mei\ss ner (1997)}
Borasoy\, B., Mei\ss ner,\, Ulf-G. (1997):
Chiral Expansion of Baryon Masses and $\sigma$-Terms.
Ann.\,Phys.\,(N.Y.) {\bf 254}, 192-232
%
\bibitem{}{bos} {Bos et al. (1997)}
Bos,\,J.W.  Chang,\,D.  Lee,\,S.C.  Lin,\,Y.C. Shih,\,H.H. (1997):
SU(3) Breaking and Baryon Magnetic Moments. 
Chin. J. Phys. (Taipei) {\bf 35},  150-155
%
\bibitem{}{CP} {Caldi and Pagels (1974)}
Caldi,\,D.G.,  Pagels\,, H. (1974):
Chiral perturbation theory and the magnetic moments of the baryon octet.
Phys. Rev. {\bf D10},  3739-3743
%
\bibitem{}{CG} {Coleman and Glashow (1961)} 
Coleman,\,S.,  Glashow\,,S.L. (1961): 
Electrodynamic properties of baryons in the unitary symmetry scheme.
Phys. Rev. Lett. {\bf 6}, 423-425
%
\bibitem{}{ecker}{Ecker (1994)} 
Ecker,\,G. (1994): Chiral invariant renormalization of the effective
pion-nucleon field theory.
Phys\,Lett. {\bf B336}, 508-517
%
\bibitem{}{juerg} {Gasser (1981)} Gasser,\,J. (1981):
Hadron Masses and the Sigma Commutator in Light of Chiral
Perturbation Theory.  Ann. Phys.\,(NY) {\bf 136}, 62-112 
%
\bibitem{}{gl85} {Gasser and Leutwyler (1985)}
Gasser,\,J., Leutwyler,\,H. (1985): Chiral perturbation theory:
Expansion in the mass of the strange quark.
Nucl. Phys. {\bf B250},  465-516
%
\bibitem{}{gss}{Gasser et al. (1988)}
Gasser,\,J.,  Sainio,\,M.E., Svarc,\,A. (1988): Nucleons
with chiral loops. Nucl.\,Phys. {\bf
B307}, 779-853
%
\bibitem{}{jm} {Jenkins and Manohar (1991)} 
Jenkins,\,E., Manohar,\,A.V. (1991): Baryon chiral perturbation theory
using a heavy fermion lagrangian. 
 Phys.\,Lett. {\bf B255}, 558-562
%
\bibitem{}{jm2} {Jenkins and Manohar (1992)} 
Jenkins,\,E., Manohar,\,A.V. (1992): The sigma term and $m_s^{3/2}$
corrections to the proton mass.
Phys.\,Lett. {\bf B281}, 336-340
%
\bibitem{}{jmr} {Jenkins and Manohar (1992)} 
Jenkins,\,E., Manohar,\,A.V. (1992) in:
{\it Effective Field Theories of the Standard Model}, ed.
Ulf-G. Mei\ss ner (World Scientific, Singapore, 1992), 113-137
%
\bibitem{}{jlms}{Jenkins et al. (1993)}
Jenkins,\,E. Luke,\,M. Manohar,\,A.V.  Savage,\, M. (1993):
Chiral perturbation theory analysis of the baryon magnetic moments. 
Phys. Lett. {\bf B302},  482-490 
%
\bibitem{}{krause} {Krause (1990)}
Krause,\,A. (1990): Baryon Matrix Elements of the Vector Current
in Chiral Perturbation Theory. Helv.\,Phys.\,Acta {\bf 63},  3-70.
%
\bibitem{}{lwd} {Leinweber et al. (1991)}
Leinweber,\,D.B.  Woloshyn,\,R.M.  Draper,\,T. (1991):
Electromagnetic structure of octet baryons.
Phys. Rev. {\bf D43},  1659-1678
%
\bibitem{}{mei:ste}{Mei\ss ner and Steininger (1997)}
Mei\ss ner,\, Ulf-G., Steininger\, S. (1997):
Baryon magnetic moments in chiral perturbation theory.
Nucl.\,Phys. {\bf B499}, 349-370
%
\bibitem{}{mue:mei}{M\"uller and Mei\ss ner (1997)}
M\"uller,\, G., Mei\ss ner,\, Ulf-G. (1997):
Renormalization of the three-flavor Lagrangian in heavy baryon
chiral perturbation theory.
Nucl.\,Phys. {\bf B492}, 379-416
%
\bibitem{}{ste:mei}{Steininger and Mei\ss ner (1997)}
Steininger\, S., Mei\ss ner,\, Ulf-G. (1997):
Threshold kaon photo- and electroproduction in SU(3) baryon
chiral perturbation theory.
Phys.\,Lett. {\bf B391}, 446-450
%
\end{thebibliography}
\end{document}